\documentclass[letterpaper,journal]{IEEEtran}
\usepackage{amsmath,amssymb}
\usepackage{subfigure}
\usepackage{graphicx,graphics,color,psfrag}
\usepackage{cite,balance} 
\usepackage[caption=false,font=normalsize,labelfont=sf,textfont=sf]{subfig}
\allowdisplaybreaks
\usepackage{accents}
\usepackage{amsthm}
\usepackage{bm}
\usepackage[english]{babel}
\usepackage{multirow}
\usepackage{enumerate}
\usepackage{cases}
\usepackage{stfloats}
\usepackage{dsfont}
\usepackage{soul}
\usepackage{amsfonts}
\usepackage{fancyhdr}
\usepackage{hhline}
\usepackage{array}
\usepackage{booktabs}
\usepackage{diagbox}
\usepackage{indentfirst}
\usepackage{url}
\usepackage{lipsum}
\usepackage{lipsum}
\usepackage[table]{xcolor}
\definecolor{pink}{rgb}{1 0.753 0.796}  
\definecolor{prim}{rgb}{0.965 0.914 0.945} 
\definecolor{aquaisland}{rgb}{0.659 0.855 0.863} 
\definecolor{whitelilac}{rgb}{0.945 0.914 0.965} 
\definecolor{ecruwhite}{rgb}{0.945 0.965 0.914}  
\definecolor{teal}{rgb}{0 0.502 0.502}  
\definecolor{aquaisland}{rgb}{0.6352 0.851 0.8078}   
\definecolor{sidecar}{rgb}{0.9686 0.9059 0.8078} 
\definecolor{peachcream}{rgb}{1 0.9451 0.8784} 
\definecolor{catskillwhite}{rgb}{0.9529 0.9765 0.9765} 
\definecolor{botticelli}{rgb}{0.8235 0.9098 0.9098} 
\definecolor{junglemist}{rgb}{0.6824 0.8314 0.8314}
\definecolor{neptune}{rgb}{0.5412 0.7530 0.7530}
\definecolor{azalea1}{rgb}{0.9686 0.7922 0.7883}
\definecolor{azalea2}{rgb}{0.9804 0.8588 0.8471}
\definecolor{whitelilac}{rgb}{0.945 0.914 0.965} 
\definecolor{prelude}{rgb}{0.8235 0.7059 0.8706}

\begin{document}
\title{Knowledge Base Enabled Semantic Communication: A Generative Perspective}
\author{Jinke Ren,~\IEEEmembership{Member,~IEEE,}~Zezhong Zhang,~\IEEEmembership{Member,~IEEE,}~Jie Xu,~\IEEEmembership{Senior Member,~IEEE,}\\~Guanying Chen,~\IEEEmembership{Member,~IEEE,}~Yaping Sun,~\IEEEmembership{Member,~IEEE,}~Ping Zhang,~\IEEEmembership{Fellow,~IEEE,}\\and Shuguang Cui,~\IEEEmembership{Fellow,~IEEE}
\thanks{J. Ren is with the Shenzhen Future Network of Intelligence Institute (FNii-Shenzhen), the School of Science and Engineering (SSE), and the Guangdong Provincial Key Laboratory of Future Networks of Intelligence, The Chinese University of Hong Kong (Shenzhen), Shenzhen 518172, China (e-mail: jinkeren@cuhk.edu.cn).}
\thanks{Z. Zhang and J. Xu are with the SSE, the FNii-Shenzhen, and the Guangdong Provincial Key Laboratory of Future Networks of Intelligence, The Chinese University of Hong Kong (Shenzhen), Shenzhen 518172, China (e-mail: zhangzezhong@cuhk.edu.cn; xujie@cuhk.edu.cn).}
\thanks{G. Chen is with the FNii-Shenzhen and the Guangdong Provincial Key Laboratory of Future Networks of Intelligence, The Chinese University of Hong Kong (Shenzhen), Shenzhen 518172, China (e-mail: chenguanying@cuhk.edu,cn).}
\thanks{Y. Sun is with the Department of Broadband Communication, Peng Cheng Laboratory, Shenzhen 518000, China, and also with the FNii-Shenzhen, The Chinese University of Hong Kong (Shenzhen), Shenzhen 518172, China (e-mail: sunyp@pcl.ac.cn).}
\thanks{P. Zhang is with the State Key Laboratory of Networking and Switching Technology, Beijing University of Posts and Telecommunications, Beijing 100876, China, and with the Department of Broadband Communication, Peng Cheng Laboratory, Shenzhen 518000, China (e-mail: pzhang@bupt.edu.cn).}
\thanks{S. Cui is with the SSE, the FNii-Shenzhen, and the Guangdong Provincial Key Laboratory of Future Networks of Intelligence, The Chinese University of Hong Kong (Shenzhen), Shenzhen 518172, China. He is also affiliated with the Department of Broadband Communication, Peng Cheng Laboratory, Shenzhen 518000, China (e-mail: shuguangcui@cuhk.edu.cn). \textit{(Corresponding author: Shuguang Cui.)} }}

\maketitle
\begin{abstract}
Semantic communication is widely touted as a key technology for propelling the sixth-generation (6G) wireless networks. However, providing effective semantic representation is quite challenging in practice. To address this issue, this article takes a crack at exploiting semantic knowledge base (KB) to usher in a new era of generative semantic communication. Via semantic KB, source messages can be characterized in low-dimensional subspaces without compromising their desired meanings, thus significantly enhancing the communication efficiency. The fundamental principle of semantic KB is first introduced, and a generative semantic communication architecture is developed by presenting three sub-KBs, namely source, task, and channel KBs. Then, the detailed construction approaches for each sub-KB are described, followed by their utilization in terms of semantic coding and transmission. A case study is also provided to showcase the superiority of generative semantic communication over conventional syntactic communication and classical semantic communication. In a nutshell, this article establishes a scientific foundation for the exciting uncharted frontier of generative semantic communication.
\end{abstract}
\section{Introduction}
\IEEEPARstart{W}ITH the unprecedented advancements in artificial intelligence (AI) and Internet of things (IoT), future sixth-generation (6G) wireless networks are experiencing a tenet shift from conventional human- and machine-type communications to ubiquitous connectivity of people, things, and intelligence. Nonetheless, classical Shannon paradigm focuses on the accurate transmission of source messages regardless of their meaning, which encounters difficulties in meeting the stringent requirements for emerging intelligent IoT applications, such as metaverse, robotics, and extended reality. This consideration gives rise to the necessity for shifting communications from ``semantic neutrality" toward ``semantic dominance" \cite{Lan:SemCom_JCIN}.

Semantic communication, originally proposed by Warren Weaver in 1953, is a revolutionary paradigm that targets the successful transmission of the desired meaning conveyed by the source \cite{Weaver:SemCom}. By extracting the so-called \textit{semantic information} from source messages and integrating task objectives into transmission, semantic communication holds great promise in significantly reducing data traffic while meeting task requirements. Due to its remarkable advantages, numerous efforts have been devoted from both academia and industry, covering key topics of semantic information theory \cite{Qin:SemCom_Arxiv}, joint source-channel coding (JSCC) \cite{Gunduz:JSCC}, and semantic-oriented radio resource management \cite{Chen:RRM}. 

Precise semantic representation is essential to reap the full potential of semantic communication. Conventionally, most prior works \cite{Qin:SemCom_Arxiv,Gunduz:JSCC,Chen:RRM} treat semantic communication in a reconstructive manner, where popular neural nets are used for extracting semantic information. However, these end-to-end designs are basically black boxes without rigorous versatility and interpretability. Some pioneering works are morphing to encompass advanced ``boosters" like knowledge graphs (KG) \cite{Zhou:Text2KG_ICC,Jiang:Text2KG_Entropy,Shi:Speech2KG_Arxiv} and feature codebooks\cite{Xie:Text2Feature_SigPro, Hu:Image2Feature_TWC, Sun:Image2Feature_Arxiv}, which supply \textit{a priori side information} to facilitate semantic coding. Nevertheless, existing studies are generally based on specified use cases with a single source modality, a particular task purpose, as well as a certain communication environment. Such dedicated designs may not generalize very well, and are much more difficult to fulfill the ambitious vision of cross-modal fusion, cross-task understanding, and cross-environment transmission in 6G.

To overcome these roadblocks, semantic knowledge base (KB) is recognized as a promising solution for effective semantic representation. Without loss of generality, semantic KB is defined as a well-structured model with powerful processing, memory, and reasoning capabilities to provide rich knowledge in supporting semantic coding and transmission. By making concession to computational overhead, semantic KB is foreseen to define a compact search space that standardizes the path for semantic representation, thereby improving the communication efficiency over classical semantic communication. In general, a semantic KB comprises three components, including source KB, task KB, and channel KB, which are beneficial to cope with various source modalities, diverse task requirements, and dynamic communication environments, respectively. 

Semantic KB can be seen as a ``catalyst" that reinvigorates the research in semantic communication, whereas its construction from scratch is typically a substantial endeavor. Fortunately, the recent astonishing success of large language models, especially the bellwether --- ChatGPT and its variants suggest an opportunity to exploit generative AI (GAI) in building semantic KB. In particular, by leveraging the superiority in pattern creations, human-like responses, and self-learning abilities, GAI-empowered semantic KB is expected to undertake a diverse array of tasks and consequently opens up a new research area of generative semantic communication. In contrast to the mission of accurately \underline{\textit{recovering}} data from noisy signals in traditional syntactic communication, generative semantic communication seeks to \underline{\textit{generate}} task-oriented contents based on received signals and GAI-empowered semantic KB. Hence, it holds great potential to significantly enhance the creativity and flexibility of future 6G wireless networks.

This article aims to provide a systematic overview on generative semantic communication. First, the fundamental principle of semantic KB is introduced along with a comprehensive review of conventional semantic KBs in the literature. A generative semantic communication system is then crystallized, wherein the entire semantic KB is divided into three sub-KBs from the source, task, and channel perspectives. The construction methods of the three sub-KBs as well as their utilization for semantic coding and transmission are respectively described. Finally, interesting topics of research for generative semantic communication are highlighted for future investigation.
\section{Preliminaries}
\subsection{Principle of Semantic KB}
\begin{figure}[!t]
\centering
\includegraphics[width=3.5in]{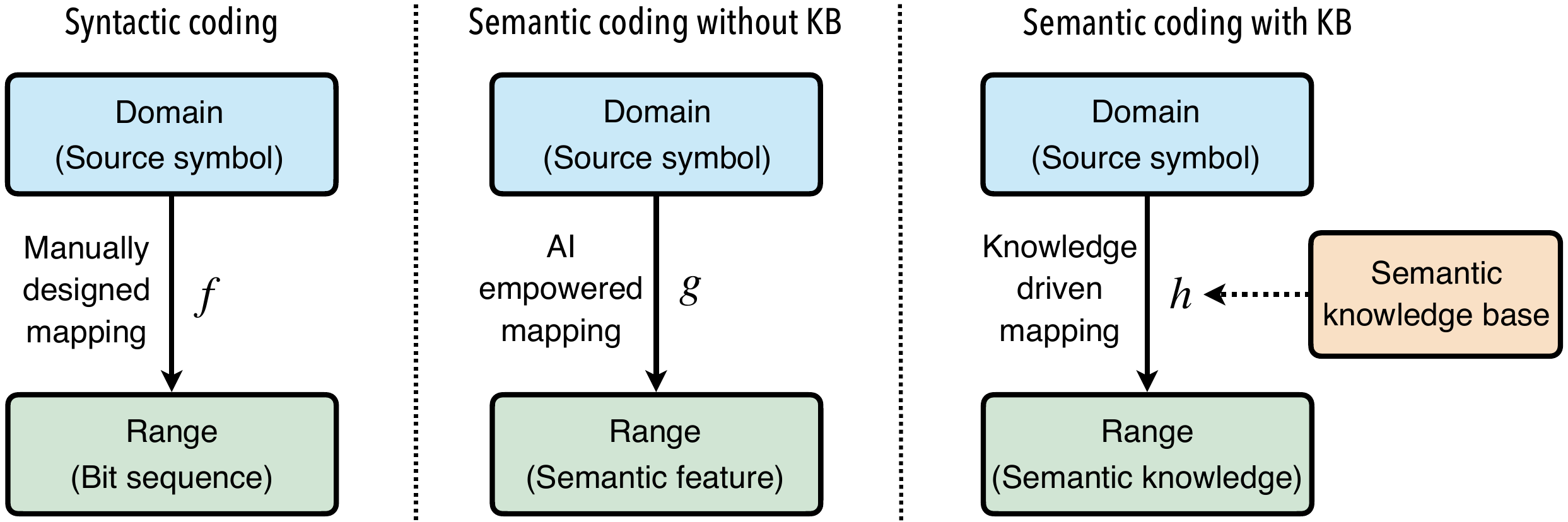}
\vspace{-1em}
\caption{A comparison of syntactic coding, semantic coding without KB, and semantic coding with KB.}
\label{fig.1}
\end{figure}

Semantic KB lies at the heart of the distinction between syntactic communication and semantic communication, especially for the coding mechanism as illustrated in Fig. \ref{fig.1}. In conventional syntactic communication, the source message is considered a series of symbols that are encoded into bit sequences before transmission. The mapping function from the domain to the range focuses on the statistical structure of the symbols and is often designed manually based on communication engineers' expertise and experience. Although already successful, the coding capacity of such ``bit-level" mechanisms is ultimately bounded by \textit{Shannon limit} because they treat all symbols as equally important.

By contrast, semantic communication only transmits essential information relevant to the task or action at the receiver rather than every symbol. As a consequence, efficiently identifying the semantic information becomes the normal objective of semantic coding. Recently, the availability of big data has motivated a substantial amount of work to use AI algorithms, particularly neural nets for extracting semantic features, thus enabling the semantic coding without KB mechanism, as shown in the middle part of Fig. \ref{fig.1}. Despite their prevalence in discovering intricate structure from source messages, such ``feature-level" mechanisms are typically short of causality since the mapping function resides in a large number of parameters of the neural nets. Perhaps more importantly, semantic coding by large-scale neural nets is often cumbersome and may occupy a considerable amount of computational resources. 

A very simple way to solve this dilemma is to reduce the neural net size. Nevertheless, small-scale nets are difficult to deal with high-dimensional data in general, and may lose some essential information, thus leading to a side effect of performance degradation. Fortunately, Wright and Ma \cite{Wright:Book} have demonstrated convincingly that the intrinsic structure of natural data typically ``lives" in a latent space with low dimensions. This critical insight inspires a new coding mechanism in which the mapping function is established with the assistance of a semantic KB, as depicted in the right half of Fig. \ref{fig.1}. \textit{The principle of the semantic KB is to define an effective search space that standardizes the path for generating knowledge from the source message}. Hence, GAI technologies are quite suitable in building semantic KB since they excel at pattern creations and possess exceptional self-learning abilities. In particular, such ``knowledge-level" mechanisms are driven by both model and data, and the encoded result is referred to as ``semantic knowledge".

\begin{table*}[t]
\centering
\caption{An overview of existing semantic knowledge base}\label{tab:overview}
\renewcommand{\arraystretch}{1.5}
\begin{tabular}{>{\raggedright\arraybackslash}p{1.4cm}|p{5.3cm}|>{\raggedright\arraybackslash}p{4.5cm}|>{\raggedright\arraybackslash}p{1cm}|>{\raggedright\arraybackslash}p{1.4cm}}
\hline
\hline
\cellcolor{prim}{\textbf{Category}} & \cellcolor{catskillwhite}{\textbf{Strength}} & \cellcolor{whitelilac}{\textbf{Weakness}} & \cellcolor{ecruwhite}{\textbf{Source}} & \cellcolor{peachcream}{\textbf{References}} \\
\hline
\multirow{2}{=}{Knowledge graph}  & \multirow{2}{=}{Structured knowledge representation\\ Remarkable scalability and flexibility\\ Powerful reasoning and querying capabilities} & \multirow{2}{=}{High maintenance cost\\No uniform standard} & Text & \cite{Zhou:Text2KG_ICC,Jiang:Text2KG_Entropy} \\
\cline{4-5}
& & &  Speech & \cite{Shi:Speech2KG_Arxiv}\\
\hline
\multirow{2}{=}{Feature codebook} & \multirow{2}{=}{Superior robustness against semantic noise\\ Lightweight storage requirement } & \multirow{2}{=}{Limited theoretical foundation\\Weak vertical generalizability\\Difficulties in designing neural nets} & Text & \cite{Xie:Text2Feature_SigPro} \\
\cline{4-5}
& & &  Image & \cite{Hu:Image2Feature_TWC, Sun:Image2Feature_Arxiv} \\
\hline
\multirow{2}{=}{Empirical dataset} & \multirow{2}{=}{Low deployment complexity\\Solid statistical interpretability} & \multirow{2}{=}{Notable data collection cost\\ Sensitive to data quality\\ Severe privacy leakage risk} & \multirow{2}{=}{Image} & \multirow{2}{=}{\cite{Letaief:ImageDataset2KB_JSAC}}\\
& & & & \\
\hline
\hline
\end{tabular}
\end{table*}
\subsection{Review of Conventional Storage-based Semantic KB}
Before proceeding to elaborate the GAI-empowered semantic KB, this section discusses three types of semantic KBs developed in recent studies, namely KG, feature codebook, and empirical dataset, as summarized in Table \ref{tab:overview}.

The terminology of KG was officially released by Google in 2012, which is also known as semantic network in the context of AI. A KG is commonly visualized as a directed labeled graph consisting of multiple nodes and edges, where a node is associated with a real-world entity (e.g., people, object, and concept) and an edge connects a pair of nodes to capture their relationship of interest (e.g., a partnership between two people). Mathematically, each knowledge element in a KG is characterized by the so-called \textit{factual triple} --- (head, relation, and tail). This nifty format allows a KG to integrate a tremendous volume of information extracted from various data sources and store them in a graph-structured manner. KGs work pretty well at organizing sequential data like text and speech, and thus can be exploited to support semantic coding by mapping a series of symbols (e.g., words) to a single knowledge element. Moreover, KGs are able to predict missing symbols and correct erroneous messages by means of reasoning and querying. Although a well-structured KG can provide abundant knowledge for semantic coding, its maintenance often requires much manual intervention. 

The feature codebook is developed to achieve further automation, which corresponds to a group of hand-engineered feature vectors with a specific form of discrete codebook. Each feature vector can be seen as a ``representative sample" in the feature space, and the whole feature codebook is likely to become a ``dictionary" that restrains the range of the mapping function in semantic coding. For example, the transmitter can associate each source symbol with a feature vector in the local codebook, where semantically related symbols end up close to each other in the feature space. Once the transmitter and the receiver share an identical codebook, it turns out that only the index of the selected feature vector needs to be exchanged, therefore bringing about a significant reduction in communication overhead and a notable improvement in resiliency against noise. 

In addition to pre-training a feature codebook beforehand, one can alternatively take the raw empirical dataset (e.g., text corpus) to facilitate semantic coding since it contains explicit statistical information of the training examples. This rather straightforward way could also improve the communication efficiency by implementing a transmitter encoder to distill the residual information of source messages beyond those in the empirical dataset. Moreover, domain adaptation technology can be used to narrow the statistical discrepancy between source messages and training examples, such that the transmitter encoder is able to generalize well far from the empirical dataset without much retraining cost.

Despite such progress, the KG, feature codebook, and empirical dataset generally serve as database, i.e., a store of knowledge, which suffers from poor versatility in practical systems with multiple source modalities, diverse task requirements, and dynamic communication environments. In other words, these designs have to be tailored to specific use cases and do not appear well suited to out-of-domain or context, which probably resists the sought-after goal of cross-modal fusion, cross-task understanding, and cross-environment transmission in semantic communication. Meeting such challenges resorts to new ideas, new tools, and a willingness to think outside the confines of existing studies.
\section{Generative Semantic Communication} 
\subsection{The Promise of Generative AI in Communications}\label{Mona Lisa}
The past year has witnessed significant breakthroughs in GAI, which has truly taken the world by storm and is reshaping a wide range of fields like science, business, and art. By learning patterns from a huge amount of data along with an in-depth understanding, large GAI models, such as ChatGPT, Llama, and Falcon are capable of providing valuable scholarly information, optimizing complex product designs as well as producing amazing artistic creations. Indeed, these large GAI models stand for the state-of-the-art KBs at present and are prone to be used in semantic communication. Imagine a use case wherein a transmitter aims to send a portrait painting, for example, Mona Lisa, to a distant receiver in real time. In classical semantic communication systems, the transmitter directly performs semantic and channel coding (lossy or lossless) on the original picture, and then sends the encoded result to the receiver. Instead, assuming that the transmitter and the receiver have equal access to a well-trained large GAI model like ChatGPT with unlimited bandwidth, while Mona Lisa has been completely parsed therein. In such circumstances, the transmitter would prefer to send its dialogue with ChatGPT, say ``[prompt]: produce the famous painting --- Mona Lisa" to the receiver. Accordingly, the receiver generates the desired result by invoking ChatGPT through the prompt. Other information about Mona Lisa, such as its author and history, can also be provided with additional prompts depending on user aspirations and appetite.
\begin{figure*}[t]
\centering
\includegraphics[width=6in]{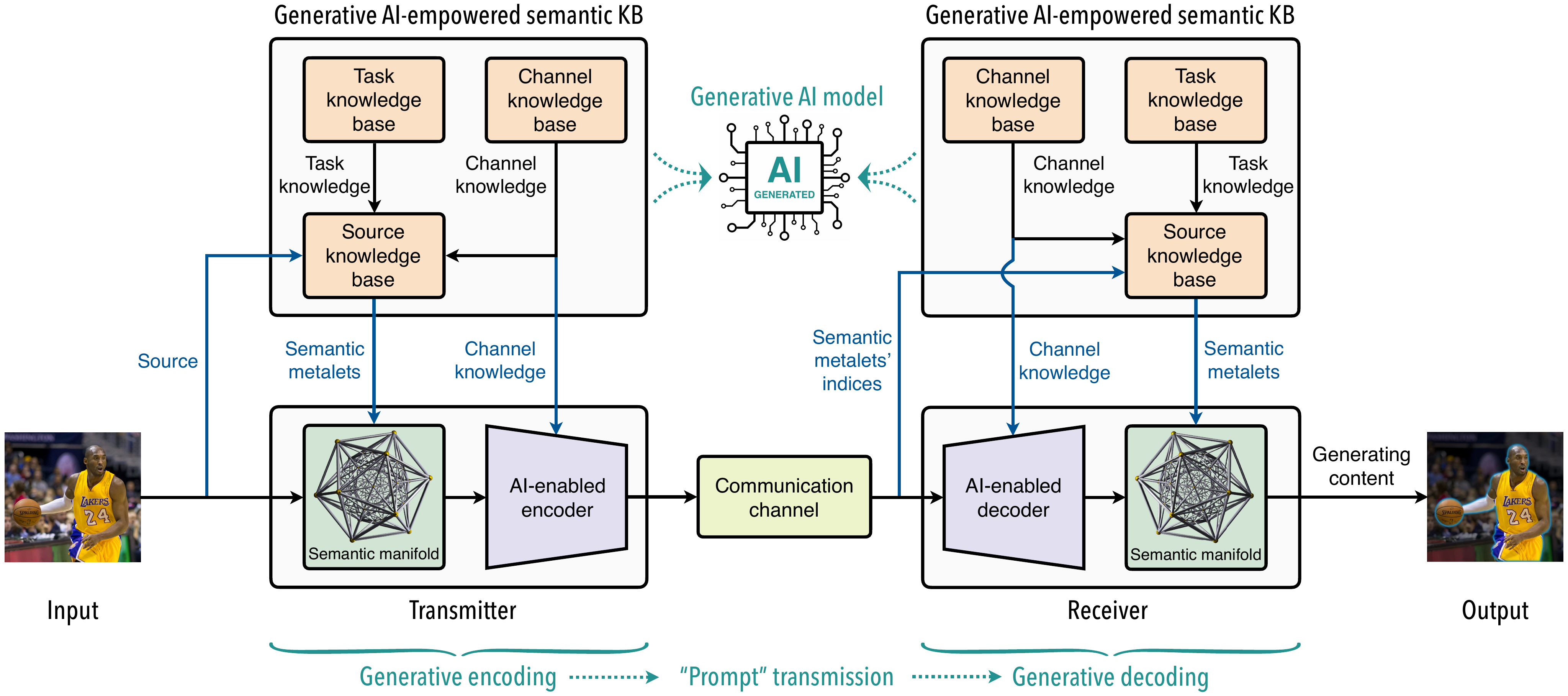}
\caption{Generic schematic diagram of a generative semantic communication system.}
\label{fig.2}
\end{figure*}
\subsection{KB-enabled Generative Semantic Communication}
As can be observed, the above use case does not consider the practical deployment of large GAI models in resource-constrained devices, but its prospect reinforces a transformative trend of embracing GAI in communications. Recent years have also seen several on-device GAI models such as Google Gemini Nano-2\footnote{https://blog.google/technology/ai/google-gemini-ai/\#sundar-note} and Microsoft Phi-2\footnote{https://www.microsoft.com/en-us/research/blog/phi-2-the-surprising-power-of-small-language-models}. However, building a generative semantic communication system is still a major undertaking, to the extent that this article makes an initial attempt at the semantic KB by integrating popular GAI models and classical storage-based KBs. By functional, a semantic KB could be divided into three parts, including source KB, task KB, and channel KB, as described below.
\begin{itemize}
\item[$\vartriangleright$] 
\textit{Source KB} is the ``kernel" of the entire semantic KB, which possesses strong generative capabilities and is akin to the large GAI models as outlined in Section \ref{Mona Lisa}. Its inputs include source messages, task knowledge, and channel knowledge, where source messages come in multiple modalities (e.g., text, speech, and image), while the latter two are provided by task KB and channel KB, respectively. Its output is a sorted group of metamodels (e.g., embedding vector and neural net) that take the helm of semantic representation. In particular, the metamodel is reminiscent of the concept of wavelet in signal processing and is henceforth named as \textit{semantic metalet}. 
\item[$\vartriangleright$] 
\textit{Task KB} can be seen as an ``assistant" to the source KB. Its input is typically a complicated task command in natural language and its output is some concise task knowledge in the form of discrete tokens. An abstract view of task KB, that frees it from any particular instantiation, is that it breaks an intricate task into several simple sub-tasks in an automatic loop. By implementing so, the transmitter will get a grokking toward the task requirement without a need for human oversight. 
\item[$\vartriangleright$]
\textit{Channel KB} is perceived as a ``mirror" that is designed to reflect the communication environment, such as the locations of obstacles and scatterers, as well as the configurations of intelligent reflecting surfaces. In general, its inputs are the geographical locations of the transmitter and the receiver. Its output is a wealth of channel knowledge (e.g., channel state information (CSI) and signal-to-noise ratio (SNR)) that are collected to empower communication designs like channel coding.
\end{itemize} 

To illustrate the way these three sub-KBs team up, we consider a point-to-point communication system as shown in Fig. \ref{fig.2}. A transmitter has a high-resolution image, for example, Kobe Bryant, while a receiver attempts to separate his portrait from the background. In accordance with the ``Mona Lisa" case introduced earlier, the transmitter will first invoke its task KB and channel KB to acquire a thorough understanding of the task requirement and channel condition. Then, the generated task knowledge and channel knowledge facilitate the source KB to judiciously activate a group of semantic metalets, which are customized to object segmentation and fit the communication environment as well. These semantic metalets carve out a low-dimensional subspace termed \textit{semantic manifold}, wherein an AI-enabled encoder encodes the original image in conjunction with the channel KB. The encoded result and the indices of the semantic metalets are then embedded into a data packet,  which is sent to the receiver through a communication channel. Likewise, the receiver identifies the indices within the data packet, followed by establishing an identical semantic manifold and generating the portrait (Kobe Bryant) based on the decoded result. Since the data packet instructs the receiver to generate the desired content, it is an analogy to the ``prompt" for large GAI models. Accordingly, the coding processes at the transmitter and the receiver are referred to as \textit{generative encoding} and \textit{generative decoding}, respectively. In particular, the receiver shall also invoke its task KB to calibrate erroneous indices caused by channel impairments (e.g., channel fading, interference, and noise).  

In short, source KB partners with task and channel KBs to abstract away a compact topology for semantic representation, thus improving the coding efficiency. Channel KB further provides accurate channel information to reshape the semantic knowledge in conformity with the communication environment. All three sub-KBs can be constructed by GAI models, and their seamless integration becomes a cornerstone for generative semantic communication.
\subsection{Key Metrics of Semantic KB}
The performance of a generative semantic communication system is highly dependent on the quality of its semantic KB. Therefore, this article proposes three key metrics toward semantic KB as follows.
\begin{itemize}
\item[$\vartriangleright$] \textit{Efficacy} refers to the gains that a semantic KB can provide to the communication performance, which is evaluated from two perspectives --- accuracy and efficiency. Specifically, accuracy measures the semantic similarity between the generated content and the desired result (e.g., peak signal-to-noise ratio (PSNR) in image restoration). Efficiency quantifies the cost of deploying a semantic KB, including processing latency, energy consumption, storage resources, etc. 
\item[$\vartriangleright$] \textit{Capacity} is defined as the total size of the semantic space sculpted by a semantic KB, in which source messages can be reliably encoded/decoded without unwarranted information loss. Capacity reflects the representation ability of a semantic KB and is closely related to the quality and quantity of semantic metalets in general. Oftentimes, more semantic metalets imply larger capacity. 
\item[$\vartriangleright$] \textit{Scalability} characterizes the degree to which a semantic KB organizes, revises, and updates itself in response to the possible changes inherent in semantic communication, such as task variations and channel dynamics. Typically, a well-scalable semantic KB is able to carry out stable evolution and withstand potential knowledge overlap and logical contradictions.
\end{itemize} 
\begin{figure}[!t]
\centering
\includegraphics[width=3.5in]{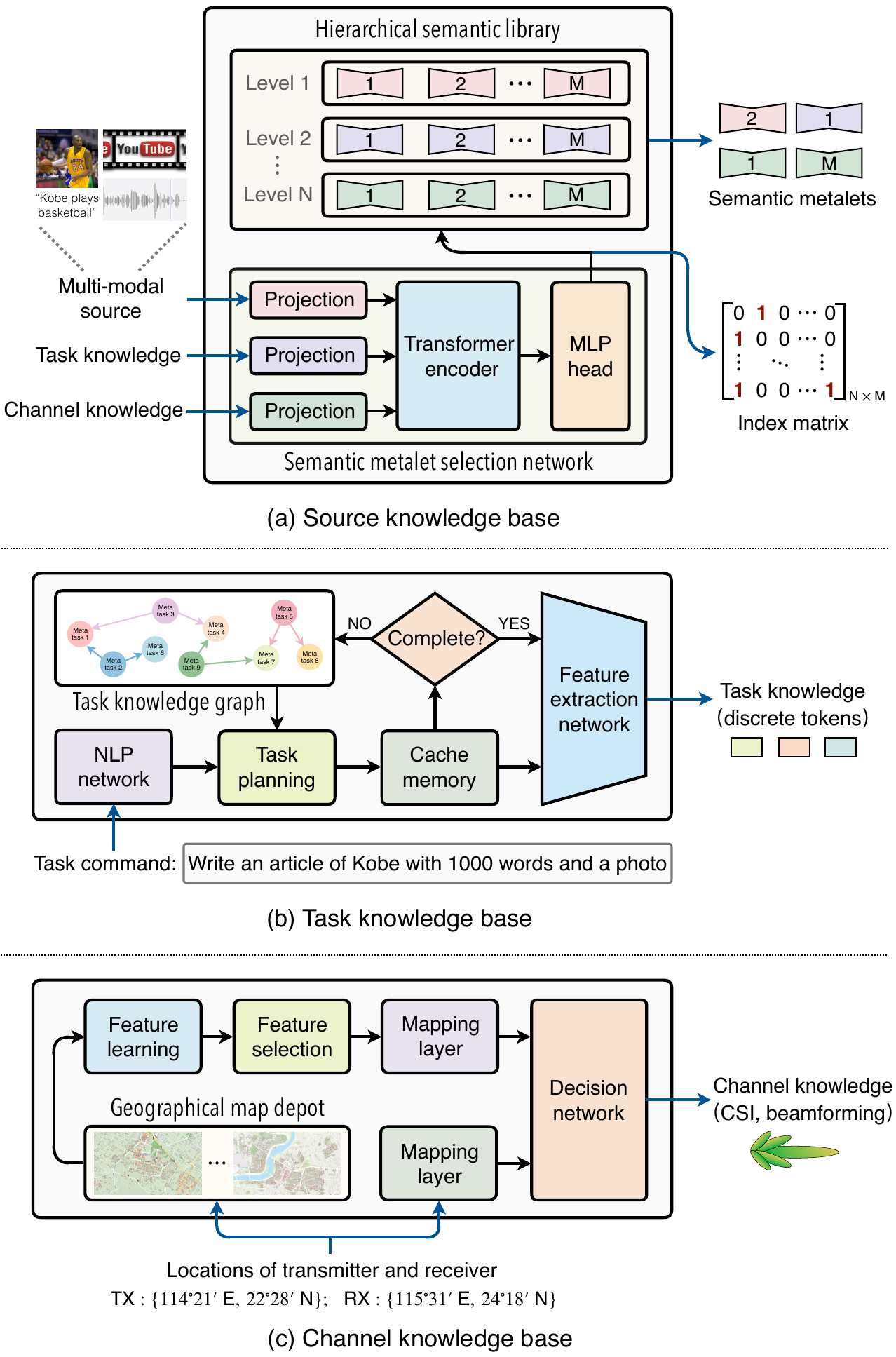}
\caption{Conceptual structures of the three sub-KBs.}
\label{fig.3}
\end{figure}
\section{Constructing Semantic KB}
\subsection{Source KB}
As highlighted, source KB strives to offer an effective way for semantic representation. It is normally composed of two modules, i.e., hierarchical semantic library and semantic metalet selection network, as shown in Fig. \ref{fig.3}(a).

Hierarchical semantic library is a collection of semantic metalets with multiple levels of abstraction, where each semantic metalet characterizes one type of data attribute, such as image texture and audio spectrogram. High-level semantic metalets are much coarser than low-level ones in general. Different levels of semantic metalets are mutually complementary to accomplish various tasks. Such an organization works well for natural data that are often compositional hierarchies. For example, in images, local connections of edges sketch motifs, motifs assemble into parts, and parts constitute objects. So far there have been two promising candidates for semantic metalets. One is \textit{embedding vector}, which is offline trained ahead of time using a huge amount of data with respect to a specific objective like image reconstruction. The other is \textit{shallow neural net} that could extract intrinsic features from multi-modal data by flexible combination. In particular, a recent work \cite{Esser:Tramsformer2modular_CVPR} has yielded surprising results that 16,384 vectors are likely to achieve high-resolution image synthesis, which suggests the possibility of using embedding vectors as semantic metalets. Moreover, the semantic library is sustainable since the number of semantic metalets may grow with the cumulative contextual knowledge.

Semantic metalet selection network serves as a ``catalog” that identifies the desirable semantic metalets with joint source, task, and channel awareness. As an instance, for image transmission under a poor channel condition, the selection network would prioritize coarse-grained semantic metalets tailored to images over fine-grained ones. Due to the stunning success of GAI models, especially the \textit{transformer} in processing multi-modal data, this article presents a simple yet effective selection network, as depicted in the bottom half of  Fig. \ref{fig.3}(a). It consists of multiple modality-specific projection layers, a transformer encoder, and a multilayer perceptron (MLP) head. Each projection layer cuts the input data into pieces (e.g., image patches), shapes each linearly, and prepends position and learnable tokens to produce a sequence of embeddings. Then the transformer encoder captures the long-range correlation of these embeddings and extracts structural information using the self-attention mechanism. The output is further passed to the MLP head, which rules out task-irrelevant information and returns a mapping matrix to suggest the indices of the 
selected semantic metalets. The selection network and the semantic library can be collectively trained, while some off-the-shelf pre-trained models might be used almost out of the box.
\subsection{Task KB}
A holy grail of task KB is the thorough understanding of task requirements, which basically shares the same spirit as human brains, hence inspiring the top-down design shown in Fig. \ref{fig.3}(b). As observed, it includes six key components: a natural language processing (NLP) network (e.g., recurrent neural net), a task planning module, a cache memory, a decision module, a task KG, and a feature extraction network (e.g., graph neural net). In particular, the task KG comprises multiple nodes and edges, where every node stands for an executable meta-task (e.g., write articles, draw paintings, and debug codes) and each edge defines the relationship between two nodes (e.g., causality, symbiosis, and progressivity). 

Let us consider the instance in Fig. \ref{fig.3}(b) to briefly describe the working mechanism of the task KB. First, the NLP network decomposes the given task command into a set of micro-commands, i.e., write an article, fix the number of words = 1000, provide a photo, and ensure the protagonist is Kobe. Then, the task planning network takes these micro-commands as input and maps out a connected subgraph by navigating over the task KG. The output is subsequently backed up in the cache memory, while a copy is delivered to the decision module as well. If the subgraph is factually rational and fulfills the task requirement, it will be transferred to the feature extraction network. Otherwise, the planning network is aroused to reorganize a new subgraph and overwrite the previous one in the cache memory. This feedback loop is repeated several times until an expected subgraph is created that fully interprets the original task command. Eventually, the feature extraction network condenses task knowledge from the final subgraph and presents it in a form that source KB can use to select semantic metalets. Indeed, this \textit{recursion of thought} endows the task KB with powerful reasoning and error-correction abilities, making it easy to produce explicit execution steps with little inductive bias or logical ambiguity. Notably, the task KG needs perpetual refinement, steered by the dual compass of graph theory and empirical experiments.
\subsection{Channel KB}
The ultimate mission of channel KB is to provide real-time channel information with minuscule communication overhead, which can be realized by the chain structure illustrated in Fig. 3(c). It is composed of six building blocks: a geographical map depot, a feature learning network (e.g., SegFormer or U-Net), a feature selection module, two mapping layers (e.g., fully-connected layer), and a decision network (e.g., fully convolutional net). A multitude of high-definition (HD) maps, either 2D or 3D, are stored in the depot, which contain rich details not normally available on traditional maps. In practice, these HD maps are usually constructed via satellite imagery or radar sensing. 

In contrast to classical training-based channel estimation, channel KB harnesses the endogenous knowledge involved in physical environments to realize light-training or even training-free channel acquisition. Specifically, the transmitter and the receiver first determine their respective locations using prevalent localization techniques, such as the global positioning system and BeiDou. Next, the location information undergoes a stack of blocks over two parallel pipelines. On one hand, a mapping layer converts it into a vector, referred to as location embedding. On the other hand, it takes as a datum to select the site-specific HD map from the geographical map depot. The resulting output is then loaded into the feature learning network, which extracts a smorgasbord of environmental features (e.g., buildings, roads, and barriers) and organizes them in a feature map. Afterward, the feature selection module identifies a part of features that are beneficial for channel acquisition, followed by a mapping layer reshaping them to a feature embedding compatible with the location embedding. Finally, the decision network integrates both embeddings and generates the desired channel knowledge accordingly. In particular, the environmental features could be collected to build channel knowledge maps using GAI models (e.g., generative adversarial net), which directly reflect the intrinsic properties of the propagation environments \cite{Zeng:CKM_Arxiv}. As such, it is highly appealing to incorporate them into channel KB, thereby obviating the need for online feature extraction.
\section{KB-Driven Semantic Coding and Transmission}
Building upon the constructed semantic KB, this section showcases how it enables generative semantic communication and presents a testament for performance evaluation.
\subsection{KB-empowered Semantic Coding}
The coding process in semantic communication typically entails two parts, i.e., semantic coding and channel coding, which are often designed independently. Semantic KB is readily amenable to such a separate coding architecture. For example, when using shallow neural nets as semantic metalets, the transmitter would systematically combine them and use the assemble net to extract the semantic information in the source message. Given the channel knowledge supplied by channel KB, the transmitter could insert redundant symbols to make the digital signal resistant to channel distortions. At the receiver side, it performs channel decoding and semantic decoding in an exactly symmetric way, followed by yielding a desired content in a generative manner.

Although this elegant decomposition has been fairly well recognized, there is a growing consensus that it may become obsolete in the future due to constraints on block code length. As an alternative, JSCC is a favored option that can be implemented with the aid of semantic KBs. A common direction is to employ an \textit{autoencoder} as the backbone structure. More precisely, an encoder and a decoder are deployed at the transmitter and the receiver, which are instantiated as neural nets that take charge of JSCC encoding and decoding, respectively. In particular, GAI models, such as transformer and diffusion models, can be utilized, where the former adepts in semantic extraction while the latter excels at content generation. The communication channel is viewed as a non-trainable layer between the encoder and decoder. Notably, the autoencoder design is closely related to the channel condition. For example, the output layer of the encoder should fit the channel dimensions. A simple strategy is to learn a set of autoencoders associated with different ranges of SNR and resource budget. In this way, the transmitter and the receiver could dynamically select a suitable autoencoder before transmission on top of the channel knowledge offered by channel KB. Another solution is to integrate an attention feature module, which fuses semantic features with the CSI in the training stage, making the system robust against the time-variation nature of the communication environment \cite{Gunduz:JSCC}.
\subsection{KB-assisted Semantic Transmission}
Semantic KB not only establishes the overarching principle and guidelines for semantic coding but also serves as a congenial conduit to expedite semantic transmission. Specifically, conventional channel estimation approaches rely on sophisticated pilot exchanges, which may not work well in ultra-dense networks with massive antennas because of the significant signaling overhead. By contrast, semantic KBs, especially the channel KB enable the acquisition of close-to-instantaneous CSI in a light-training or even training-free manner, therefore achieving huge savings in terms of time and spectrum resources \cite{Zeng:CKM_Arxiv}. 

On the other hand, channel KB is able to make coarse prediction about future channel conditions by mining the spatial-temporal evolution nature inherent to the communication environment. The inferred result can be further refined by online channel training with a little pilot overhead. Such forward-thinking practices permit the network controller to orchestrate effective network planning, such as link scheduling, location placement, and interference cancellation. Meanwhile, economical radio resource management (e.g., power control, bandwidth allocation, and beam selection) can be achieved by optimizing a meaningful utility function, such as system latency and energy consumption. It is noteworthy that these designs can be scaled out and up to future 6G wireless networks with multiple transmitter-receiver pairs, which enhance the possibility for building an environment-aware generative semantic communication system.
\subsection{A Case Study}
This subsection conducts an empirical exploration to validate the benefit of generative semantic communication. The default configurations are described as follows. Consider a point-to-point communication system, wherein a transmitter possesses a blurry image with 128$\times$128 pixels while a receiver aims at image restoration. An additive white Gaussian noise channel model is simulated in conjunction with multiple levels of SNR\footnote{In practice, the impariments of wireless channels (e.g., channel fading and frequency offsets) can be addressed by standard channel equalization techniques.}. Since the task requirement and channel model are quite clear, we focus on the utilization of source KB, which comprises 1024 embedding vectors serving as semantic metalets and is shared by both the transmitter and the receiver. The number of selected embedding vectors is set to 512. A convolutional neural net with 3 self-attention blocks and 14 residual blocks is employed as the transmitter encoder, which draws out the residual information from the input image not captured by the selected embedding vectors and sculpts it to fit the channel condition. Correspondingly, the receiver decoder is instantiated as a convolutional neural net with 3 self-attention blocks and 14 residual blocks, which performs decoding by merging the residual information and the selected embedding vectors. The overall architecture is trained in an end-to-end fashion under an organic supervision of large-scale data annotations. 
\begin{figure}[!t]
\centering
\includegraphics[width=3.5in]{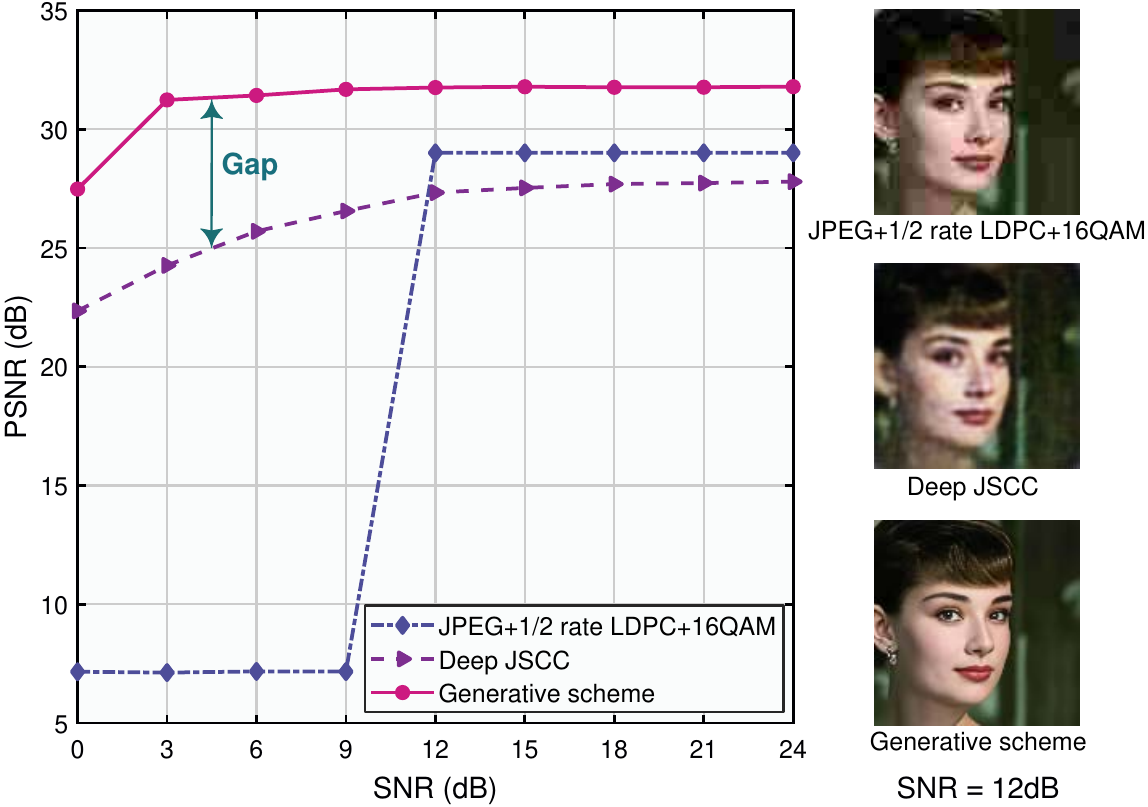}
\caption{PSNR comparison and image visualization for different schemes.}
\label{fig.4}
\end{figure}

For comparison purposes, two well-studied schemes are implemented for benchmarking, namely JPEG+1/2 rate LDPC+16QAM and Deep JSCC \cite{Gunduz:JSCC}. All schemes are evaluated on a performance metric called PSNR, which quantifies the similarity between the reconstructed image and the desired ones. A higher PSNR score indicates better performance. As shown in Fig. \ref{fig.4}, the generative scheme outperforms other two schemes across all SNR regimes, while the performance gap between the generative scheme and Deep JSCC is significant when SNR is low. A visual comparison of the reconstructed images when SNR = 12dB is presented in the right half of Fig. \ref{fig.4}, where the generative scheme produces the most pleasant result with smooth facial details. We attribute this magnificence to the utilization of semantic KB, which allows the decoder to use embedding vectors to constitute global composition, along with incorporating residual information to remedy missing motifs in a generative sense. This process echoes the forward generation step of ChatGPT, hence strengthening the creativity of generative semantic communication.
\section{Future Research Directions}
Generative semantic communication is still in the early days and many active areas of interest deserve further research.
\subsection{Consistency-aware Mechanism}
As noticed, the superiority of generative semantic communication over classical ones arises from a premise --- the transmitter and the receiver share an identical semantic KB. However, keeping their semantic KBs the same all the time is well beyond reach in practice. A partial solution is to check the consistency through frequent signaling interactions, but it may induce high communication cost. Another possibility is to develop an automatic trigger mechanism anchored on the gap between the generated content and the desired result, thus avoiding unnecessary signaling and enhancing communication efficiency. Moreover, it is likely to mimic human-to-human communication where conversations can be made as long as their local brains are relatively consistent at a semantic level.
\subsection{Multi-agent Update Strategy}
Due to the variability in wireless communications, a static semantic KB may suffer from performance degradation after a long period of time. This necessitates the dynamic evolution of semantic KB to stay current with the latest requirements. On one hand, each type of sub-KB can update itself by emerging AI algorithms, such as contrastive learning and knowledge distillation. On the other hand, the three sub-KBs are allowed to update together since they are of close concern to each other. For example, the source data in surveillance scenarios often contain much environmental information, which can be used to update channel KB. Furthermore, it would be interesting to update semantic KBs across a multitude of devices in a distributed fashion. Federated learning is a potential paradigm, but associated designs like resource allocation and interference management require deliberate and coordinated efforts.
\subsection{Privacy and Security Protection}
The deployment of semantic KB in generative semantic communication poses a variety of risks to privacy and security that cannot be overlooked. A hidden third-party will probably identify the semantic information if it obtains the shared semantic KB for a transmitter-receiver pair. By further scrutinizing the semantic KB, it is also able to disseminate biased content during semantic transmission, and in turn leads to a failure of semantic decoding. A possible solution is to combine physical-layer security techniques (e.g., directional and spatial modulations) to prevent illegal eavesdropping. Additionally, cutting-edge algorithms and hardware, such as adversarial training and trusted execution environment can be employed to reduce system vulnerability and eliminate the long-tail effect with respect to 
malicious poisoning attacks. 
\section{Conclusions}
Semantic KB is an upcoming technology for improving communication efficiency in 6G. This article reviewed the fundamental concept of semantic KB, provided a holistic approach to constructing three types of sub-KBs, and discussed how they contribute to generative semantic communication. While the deployment of semantic KB has profound ramifications on semantic coding and transmission, substantial technical and engineering issues remain to be addressed. We hope that this article can offer a ``lighthouse method" for generative semantic communication, whereas its implementation in future communication networks is too large in scope that calls for a combination of complementary disciplines like GAI and wireless communications.
\section{Acknowledgments}
The work was supported in part by NSFC with Grant No. 62293482, the Basic Research Project No. HZQB-KCZYZ-2021067 of Hetao Shenzhen-HK S\&T Cooperation Zone, the Shenzhen Outstanding Talents Training Fund 202002, the Guangdong Research Projects No. 2017ZT07X152 and No. 2019CX01X104, the Guangdong Provincial Key Laboratory of Future Networks of Intelligence (Grant No. 2022B1212010001), and the Shenzhen Key Laboratory of Big Data and Artificial Intelligence (Grant No. ZDSYS201707251409055). This work was also supported in part by the NSFC with Grants No. U2001208 and 92267202, the NSFC with Grant No. 62301471, the Major Key Project of PCL Department of Broadband Communication (PCL2023AS1-1), the Guangdong Basic and Applied Basic Research Foundation with Grant No. 2023A1515110425, and the Shenzhen Science and Technology Program with Grant  No. KJZD20230923115104009.

\vspace{-3em}
\begin{IEEEbiographynophoto} {Jinke Ren} [Member, IEEE] 
is currently a Postdoctoral Research Fellow with the Shenzhen Future Network of Intelligence Institute (FNii-Shenzhen), the School of Science and Engineering (SSE), and the Guangdong Provincial Key Laboratory of Future Networks of Intelligence, The Chinese University of Hong Kong, Shenzhen, China. His research interests are in computer vision, embodied intelligence, blockchains, and machine learning with applications to Internet of things and wireless networks.
\end{IEEEbiographynophoto}

\begin{IEEEbiographynophoto}{Zezhong Zhang} [Member, IEEE] 
is currently a Research Assistant Professor with the School of Science and Engineering (SSE), the Shenzhen Future Network of Intelligence Institute (FNii-Shenzhen), and the Guangdong Provincial Key Laboratory of Future Networks of Intelligence, The Chinese University of Hong Kong, Shenzhen, China. His research interests are in the areas of edge learning, radio map estimation, integrated sensing and communication and B5G technologies.
\end{IEEEbiographynophoto}

\begin{IEEEbiographynophoto}{Jie Xu} [Senior Member, IEEE] 
is currently an Associate Professor (Tenured) with the School of Science and Engineering (SSE), the Shenzhen Future Network of Intelligence Institute (FNii-Shenzhen), and the Guangdong Provincial Key Laboratory of Future Networks of Intelligence, The Chinese University of Hong Kong, Shenzhen, China. His research interests include wireless communications, wireless information and power transfer, UAV communications, edge computing and intelligence, and integrated sensing and communication (ISAC).
\end{IEEEbiographynophoto}

\begin{IEEEbiographynophoto}{Guanying Chen} [Member, IEEE] 
is currently a Research Assistant Professor with the Shenzhen Future Network of Intelligence Institute (FNii-Shenzhen) and the Guangdong Provincial Key Laboratory of Future Networks of Intelligence, The Chinese University of Hong Kong, Shenzhen, China. His research interests focus on deep learning based methods for computer vision and semantic communication.
\end{IEEEbiographynophoto}

\begin{IEEEbiographynophoto}{Yaping Sun} [Member, IEEE] 
received the B.E. degree and the Ph.D. degree from Xidian University and Shanghai Jiao Tong University, in 2015 and 2020, respectively. From 2018 to 2019, she was a Visiting Scholar with University of Washington, USA. From 2020 to 2022, She was a Post-Doctoral Research Fellow with the Shenzhen Future Network of Intelligence Institute (FNii-Shenzhen), The Chinese University of Hong Kong, Shenzhen (CUHK-Shenzhen), China. She is currently an Assistant Researcher with the Department of Broadband, Peng Cheng Laboratory, Shenzhen, and also with the FNii-Shenzhen, CUHK-Shenzhen, China. Her research interests include mobile 3C networks and semantic communication.
\end{IEEEbiographynophoto}

\begin{IEEEbiographynophoto}{Ping Zhang} [Fellow, IEEE] 
is a Professor with Beijing University of Posts and Telecommunications, the Director of the State Key Laboratory of Networking and Switching Technology, the Director of the Department of Broadband Communication of Peng Cheng Laboratory, a member of IMT-2020 (5G) Experts Panel, and a member of Experts Panel for China’s 6G development. He served as a Chief Scientist of National Basic Research Program (973 Program), an expert in Information Technology Division of National High-Tech R\&D Program (863 Program), and a member of Consultant Committee on International Cooperation of NSFC. His research interests mainly focus on wireless communication.   
\end{IEEEbiographynophoto}

\begin{IEEEbiographynophoto}{Shuguang Cui} [Fellow, IEEE] 
received his Ph.D. from Stanford in 2005. He is now a X.Q. Deng Presidential Chair Professor at The Chinese University of Hong Kong, Shenzhen, China. His current research interest is data driven large-scale information analysis and system design. He was selected as the Thomson Reuters Highly Cited Researcher and listed in the Worlds’ Most Influential Scientific Minds by ScienceWatch in 2014. He was the recipient of the IEEE SP Society 2012 and ComSoc 2023 Marconi Best Paper Awards. He is an IEEE Fellow, Member of Both Royal Society of Canada and Canadian Academy of Engineering. 
\end{IEEEbiographynophoto}
\end{document}